\documentclass[%
 reprint,
 amsmath,amssymb,
 aps,
]{revtex4-1}
\usepackage{graphicx}% Include figure files
\usepackage{dcolumn}% Align table columns on decimal point
\usepackage{bm}% bold math
\usepackage{amsmath,amssymb}
\usepackage{slashed}
\usepackage{graphicx}
\usepackage{dsfont}
\usepackage{bm}
\usepackage[top=1.0in,bottom=1.0in,left=1.0in,right=1.0in]{geometry}
\usepackage[colorlinks,linkcolor=blue,citecolor=blue]{hyperref}
\usepackage{CJKutf8}

\begin{document}
\setlength{\oddsidemargin}{0.5cm}
\setlength{\topmargin}{-0.1cm}
\setlength{\textheight}{21cm}
\setlength{\textwidth}{15cm}
\newcommand{\be}{\begin{equation}}
\newcommand{\ee}{\end{equation}}
\newcommand{\bea}{\begin{eqnarray}}
\newcommand{\eea}{\end{eqnarray}}
\newcommand{\ba}{\begin{eqnarray}}
\newcommand{\ea}{\end{eqnarray}}

\newcommand{\fslash}{\hspace{-1.4ex}/\hspace{0.6ex} }
\newcommand{\Dslash}{D\hspace{-1.6ex}/\hspace{0.6ex} }
\newcommand{\Wslash}{W\hspace{-1.6ex}/\hspace{0.6ex} }
\newcommand{\pslash}{p\hspace{-1.ex}/\hspace{0.6ex} }
\newcommand{\kslash}{k\hspace{-1.ex}/\hspace{0.6ex} }
\newcommand{\underkslash}{{\underline k}\hspace{-1.ex}/\hspace{0.6ex} }
\newcommand{\epslash}{{\epsilon\hspace{-1.ex}/\hspace{0.6ex}}}
\newcommand{\partslash}{\partial\hspace{-1.6ex}/\hspace{0.6ex} }

\newcommand{\nn}{\nonumber}
\newcommand{\Tr}{\mbox{Tr}\;}
\newcommand{\tr}{\mbox{tr}\;}
\newcommand{\ket}[1]{\left|#1\right\rangle}
\newcommand{\bra}[1]{\left\langle#1\right|}
\newcommand{\rhoraket}[3]{\langle#1|#2|#3\rangle}
\newcommand{\brkt}[2]{\langle#1|#2\rangle}
\newcommand{\pdif}[2]{\frac{\partial #1}{\partial #2}}
\newcommand{\pndif}[3]{\frac{\partial^#1 #2}{\partial #3^#1}}
\newcommand{\pbm}[1]{\protect{\bm{#1}}}
\newcommand{\avg}[1]{\left\langle #1\right\rangle}
\newcommand{\vnabla}{\mathbf{\nabla}}
\newcommand{\notes}[1]{\fbox{\parbox{\columnwidth}{#1}}}
\newcommand{\pair}{\raisebox{-7pt}{\includegraphics[height=20pt]{pair0.pdf}}}
\newcommand{\paircrs}{\raisebox{-7pt}{\includegraphics[height=20pt]{pair0cross.pdf}}}
\newcommand{\paircc}{\raisebox{-7pt}{\includegraphics[height=20pt]{pair0cc.pdf}}}
\newcommand{\paircrscc}{\raisebox{-7pt}{\includegraphics[height=20pt]{pair0crosscc.pdf}}}
\newcommand{\pairloop}{\raisebox{-7pt}{\includegraphics[height=20pt]{pairloop.pdf}}}
\newcommand{\pairloopf}{\raisebox{-7pt}{\includegraphics[height=20pt]{pairloop4.pdf}}}
\newcommand{\pairlooph}{\raisebox{-7pt}{\includegraphics[height=20pt]{pair2looph.pdf}}}

%\preprint{APS/123-QED}

\title{Universality of Koba-Nielsen-Olesen scaling\\ in QCD at high energy and entanglement}% Force line breaks with \\

\author{Yizhuang Liu}
 \affiliation{Institute of Theoretical Physics, Jagiellonian  University,
 30-348 Krak\'{o}w, Poland.}
 \email{yizhuang.liu@uj.edu.pl}%Lines break automatically or can be forced with \\
\author{Maciej A. Nowak}%
 \email{maciej.a.nowak@uj.edu.pl}
\affiliation{
 Institute of Theoretical Physics and Mark Kac Center for Complex Systems Research, Jagiellonian University, 30-348 Krak\'{o}w, Poland
}

\author{Ismail Zahed }
\email{ismail.zahed@stonybrook.edu}
\affiliation{Center for Nuclear Theory, Department of Physics and Astronomy, Stony Brook University, Stony Brook, New York 11794--3800, USA}
%
%\affiliation{
 %Third institution, the second for Charlie Author
%
%^\author{Delta Author}
%\affiliation{%


\date{\today}% It is always \today, today,

\begin{abstract}
Using Mueller's dipole formalism for deep inelastic scattering in QCD, we formulate and solve the evolution for the generating function for the
 multiplicities of the produced particles, in hadronic processes at high energy.  The solution for the multiplicities satisfies Koba-Nielsen-Olesen (KNO) scaling, with good agreement with the recently re-analyzed data from the H1 experiment at HERA (DESY),  and the old ALEPH data for hadronic $Z$ decay at LEP (CERN).  The same scaling function with KNO scaling, carries to the hadronic multiplicities from jets in electron-positron annihilation. This agreement is {\it a priori} puzzling, since in Mueller's dipole evolution,  one accounts for virtual dipoles in a wave function, whereas in electron-positron annihilation, one describes  cross-sections of real particles. We explain the origin  of this  similarity, pointing at a particular duality between the two processes. Finally, we interpret our results from the point of view  of quantum entanglement between slow and fast degrees of freedom in QCD, and derive the  entanglement entropy pertinent to electron-positron annihilation into hadronic  jets.

\end{abstract}

\maketitle

%\tableofcontents

%\section{}
\noindent
{\bf 1.} Universality is a powerful concept permeating several branches of physics, whereby different physical systems can exhibit similar  behavior. This is usually captured  by  universal exponents, given general assumptions. Perhaps, the best example
is the universality of the critical exponents in scaling laws in the vicinity of phase changes. 
Scaling laws, {\it per se}, form an important theoretical corpus in physics. In general, they describe the functional relationship between two physical quantities, that scale with each other over a significant interval. 

In the context of high energy particle physics, the so-called Koba-Nielsen-Olesen  scaling (named KNO scaling  hereafter), formulated half a century ago, is of  paramount importance in the empirical analysis of many high energy hadronic multiplicities. Yet, it is usually  challenging  to derive from first principles in QCD.
 Historically, KNO scaling was first formulated in two, independent theoretical works
~\cite{POLYAKOV} and \cite{KNO}, which  suggested that at high energies with large Mandelstam $s$
(squared invariant mass),
the probability distribution of producing $n$ particles in a specific collision process,  scales as 
\begin{eqnarray}
\overline{n} (s) p_n(s)= f(z)
\end{eqnarray}
where $\overline{n}(s)$ is the average multiplicity at large $s$, and $z\equiv\frac{n}{\overline{n}(s)}$ is the argument of the {\it scaling function} $f(z)$.
Remarkaby,  the  KNO scaling hypothesis precedes the  emergence  of Quantum  Chromodynamics (QCD), and the advent of high energy and luminosity data currently available at  colliders.

This letter is motivated by the  recent work in~\cite{DESY}, where the DIS data from the H1 experiment at DESY were re-analyzed, with interest in an assessment  of the quantum entanglement in high energy particle physics. Clearly, the data analysed, especially for the highest energy range, shows KNO scaling (see Fig.~1). Also, the Shannon entropy of the multiplicities presented, bears some similarity to the entanglement entropy. However, the explicit form of the scaling function was unknown, and the QCD understanding of the hypothetical entanglement was not specified.

In the first  part of this letter, we discuss the unexpected {\it a priori}  fact, that identical differential equations, such  as the ones we derived  recently in~\cite{Liu:2022bru}, yield  a scaling function  that applies
to different settings at high energy, e.g.  deep inelastic scattering (DIS) and jets in electron-positron or $e^+e^-$ annihilation.  
Exploiting  the formalism based on the Banfi-Marchesini-Smye (BMS) construction~\cite{BMS},  we point to the fact that the pertinent  generating functions for the multiplicity probabilities  in the case of  DIS and jets, respectively,    are  mathematically {\it identical}, leading to similar  equations. Furthermore, in  the double logarithm approximation (DLA), we arrive at the final differential equation for the KNO function, which we then solve using methods based on analyticity. 
The resulting, parameter-free curve, does not only agree well with DIS data from the H1 experiment at DESY, and the $Z$-decay data from ALEPH at CERN, but also represents an exact prediction for future DIS experiments, alike  EIC or EicC.

In sum, this letter consists of three new results: 1/ The derivation of the KNO scaling function in QCD for both DIS and jets in the DLA; 2/ The use of the KNO scaling function in the DLA,  to show the universality of the hadronic multiplicities from current colliders, for both DIS and jets; 3/ The explicit derivation of the entanglement entropy for $e^+e^-$ annihilation into hadronic jets, to be measured at collider energies.
\\ 
\\
\noindent
{\bf 2.}
The BMS equation~\cite{BMS,Marchesini:2003nh,Caron-Huot:2015bja}, describes the ``non-global'' logarithms in the $e^+e^-$ annihilation process, and is based on the universal features of the soft divergences in the $n$-gluon contribution to the total cross section $\sigma_n$. To leading logarithm accuracy, the ``most singular'' part of $\sigma_n$,  can be effectively generated through a Markov process.  Defining the directions  of the quark-antiquark pair  as $p$ and $n$, soft gluons ($k$) are emitted from harder ones ($p$) through a universal eikonal current $\frac{gp^{\mu}}{p\cdot k}$, and the emissions are strongly ordered in time and energy. As a result, the emission depends only on the color-charges that are already present in the final state, but not on the history of how they are emitted. In the large number of colors $N_c$, the generating functional for $\sigma_n$, 
\bea
Z\big(\frac{E}{E_0};n,p;\lambda\big)=\sum_{n=0}^{\infty}\lambda^n \sigma_n
\eea
satisfies a closed integral equation~\cite{Liu:2022bru}
%~\cite{BMS,Marchesini:2003nh,Caron-Huot:2015bja}
\begin{align}\label{eq:bms}
&Z(\frac{E}{E_0};n,p;\lambda)=e^{-\bar \alpha_s \ln \frac{E}{E_0}\int d\Omega_k K(\hat k; \hat p, \hat n)} \nonumber \\
&+\bar \alpha_s \lambda \int_{E_0}^{E} \frac{d\omega}{\omega} e^{-\bar \alpha_s\ln \frac{E}{\omega} \int d\Omega_k  K(\hat k; \hat p, \hat n)} \int d\Omega_k \nonumber \\ 
&\times K(\hat k; \hat p, \hat n)Z(\frac{\omega}{E_0};n,\hat k;\lambda)Z(\frac{\omega}{E_0};p,\hat k;\lambda) \ ,
\end{align}
with the eikonalized gluonic emission kernel $K(\hat k;\hat p,\hat n)=\frac{1}{4\pi}\frac{p\cdot n}{\hat k\cdot p \hat k\cdot n}$ and $\bar{\alpha_s}=N_c \alpha_s/\pi$.
The first term is  the Sudakov contribution where all the soft gluons are virtual, and the second term is the contribution where at least one soft gluon is real. (\ref{eq:bms}) is the integral form of the BMS equation, which can brought to the  standard form discussed in~\cite{BMS,Marchesini:2003nh,Caron-Huot:2015bja}, by taking a derivative with respect to $\ln \frac{E}{E_0}$ and some re-arrangements.\\
%\frac{d}{d\ln \frac{E}{E_0}}Z(\frac{E}{E_0};n,p;\lambda)=\bar \alpha_s \int d\Omega_k \frac{\hat p\cdot \hat n}{\hat k\cdot \hat p\hat k \cdot \hat n} \bigg(-Z(\frac{E}{E_0};n,p;\lambda)+\lambda Z(\frac{E}{E_0};n,\hat k;\lambda)Z(\frac{E}{E_0};p,\hat k;\lambda)\bigg) \ .
%\end{align}
%This is the standard form of the BMS equation.
\\
\\
\noindent
{\bf 3.}  Mueller~\cite{DIPOLES} has shown that the small-$x$ evolution equations such as BFKL~\cite{BFKLa,BFKLb,BFKLc}, and BK~\cite{BKa,BKb},  are also based on a very similar branching process, where small-$x$ virtual gluons are released into the light-front wave functions (LFWFs). The same reasoning yields an evolution equation, this time for the generating function for the   distribution 
of the virtual dipoles or $Z(b,\frac{x}{x_{\rm min}},\lambda)$, which is  exactly of the form (\ref{eq:bms}), with the substitution  $\int d\Omega_k \rightarrow \int d^2b_2$. Here $b_2$ denotes the transverse position of the emitted soft gluon,  with  a different eikonalized emission kernel $K(b_2;b_0,b_1)$ that  on the transverse dipole positions. 

In fact, one can show that in the leading order,  the BMS and BK equations  map onto each through a pertinent conformal transformation~\cite{Cornabla:2007fs,Vladimirov:2016dll,STEREO}, where the asymptotic real soft gluons at $t=\infty$, map onto the virtual gluons present at $x^+=0$~\cite{Liu:2022bru}. In this sense, the mapping is a ``virtual-real'' duality, in addition to the standard interpretation that it maps rapidity divergence to UV divergence~\cite{Vladimirov:2016dll}\footnote{In non-conformal theory, the exact mapping breaks at two-loop already. But for the virtual part, it can be generalized to all orders.}. In Table~I we have highlighted this duality through a parallel between the  two constructions.  %We mention here, that the similarity between the jet physics and BK/BFKL formalism has been noted in the past~\cite{SIMILARITY1,SIMILARITY2,SIMILARITY3}, although, as far as we know,  not in the form of mathematically identical equations for corresponding generating functions. \\
%\begin{align}
\\
\\
\noindent
{\bf 4.} The BK (BMS)  equation  resums {\it single} logarithms in rapidity (energy).  However, in both cases there are two types of divergences instead of one: in the Mueller's dipole construction for the LFWFs~\cite{DIPOLES}, there are UV divergences %(in $p_n$)
when $k_\perp$ becomes large, while in the BMS construction with Wilson-line cusps,  there are rapidity-divergences when the emissions become collinear to the Wilson-lines. It is natural to resum  the double logarithms for  both. One can then show that the double logarithm approximation
(DLA) for BK (BMS) is generated from the same branching process, with one more strong ordering in dipole sizes (emitting angles) (see Table~I). The strong ordering in virtuality is preserved by the DLA limit.  Clearly, the two DLA  and the underlying size (angle)  orderings, simply map onto each other under a conformal transformation.%~\cite{KNOlarge}.
The similarity of the branching process follows from the non-Abelian three-gluon coupling  and large $N_c$ in QCD, giving rise to a binary tree branching.
\\
\\
\noindent
{\bf 5.} A unique feature of the DLA  is that the distribution of dipoles (soft gluons)  has a nontrivial KNO scaling function $f(z)$, which coincides with that suggested  many years ago for jets in~\cite{OLDJETKNO1a,OLDJETKNO2,OLDJETKNO3}. This is due to the same strong ordering in emitting angles in both cases, with Table~I making now this plausible.

\begin{table*}
\caption{\label{tab:comparasion} Mueller hierarchy and BMS hierarchy}
\begin{tabular}{ c|c|c}
\hline
& Dipole& Cusp \\
\hline\hline
distribution in& {\it virtual} gluon in wave function & {\it real} gluon in asymptotic state\\
\hline
large $N_c$& yes&yes  \\
\hline
kernel&$\frac{b^2_{10}}{b^2_{12}b^2_{20}}$& $\frac{n\cdot p}{\hat k\cdot n \hat k\cdot p}$ \\
\hline
virtual part& TMD soft factor& Sudakov form factor  \\
\hline
time ordering& in LF time&in CM time  \\
\hline
momentum ordering&decreasing $k^+$& decreasing energy $\omega$ \\
\hline
virtuality ordering& increasing&decreasing   \\
\hline
Markov process &yes&yes  \\
\hline
DLA & $b_{10}\gg b_{12} \gg...$& $\theta_{01}\gg \theta_{12}\gg....$    \\
\hline
\end{tabular}
\end{table*}

In the DLA, the equation for the generating function (\ref{eq:bms}) simplifies
\bea
\label{PB}
    Z(\rho)=e^{-\rho} +\rho\!\!\int_0^1 \!\!\!\!dx\int_0^1\!\!\!\!dy e^{-\rho(1-y)}Z(\rho xy)Z(\rho y)  \nonumber
\eea
where 
$$\rho =\frac{2C_F}{\pi \beta_0} y \ln \ln \frac{Q^2}{M^2}$$ 
with $2C_F\sim N_c$ and $\beta_0=\frac{11N_c}{12\pi}$ for large $N_c$. Defining $Z=\exp W$,  and introducing $u=2\sqrt{\rho}$, we arrive at the final equation $\Delta_2 W=e^W-1$,
where $\Delta_2$ is a radial part of the 2-dimensional Laplacian. 
This equation is reminiscent of the Poisson-Boltzmann equation, if $\Delta_2$ was the full Laplacian. For large energies, it  reduces to  
\bea
\label{PBX}
\frac{d^2W}{du^2}=e^W-1 \ .
\eea
The solution of this equation, encodes the shape of the KNO scaling function $f(z)$, through a Fourier-Laplace transform
\bea
\label{PBX1}
Z(t=-e^u)=\int_0^\infty dz\, e^{-tz}\,f(z)
\eea
A detailed investigation of (\ref{PBX}-\ref{PBX1}) can be found in our recent analysis~\cite{Liu:2022bru}. With the help of complex-analytic method, we are able to unravel the scaling function $f(z)$ both in the asymptotic region analytically, and throughout  numerically. For $z<0.1$, the scaling function can be made explicit
\bea
f(z) \sim \frac{\alpha}{z^2}\ln \frac{\alpha}{z}\exp\bigg(-\frac{1}{2}\ln^2 \frac{\alpha}{z}\ln \frac{\alpha}{z}+{\cal O}(1)\bigg) \ .\nonumber\\
\label{ysmally}
\eea
where $\alpha= 1.50972$. 
%The speed of growth is much slower than the $0+1$D reduction~\cite{DIPOLES}, but much faster than the $1+2$D case~\cite{OUR2}. 
For $z>2.0$, the scaling function behaves asymptotically as 
\bea
\label{PBXX}
f(z)=2r\bigg(rz-1+{\cal O}\bigg(\frac{\ln z}{z}\bigg)\bigg)e^{-rz} \ , \ z \rightarrow \infty \  , \nonumber\\\label{eq:fsamylarge}
\eea
where $r=2.55297$.

Since this scaling  function represents  a parameter-free QCD prediction for future experiments, including the EIC and EicC, we record our  numerical solution in  Table~II, following from  our analysis in~\cite{Liu:2022bru}. 
For completeness, we note that a moment reconstruction of $f(z)$ using different arguments, was used in the context of jets in~\cite{OLDJETKNO1a,OLDJETKNO3,BASSETTO}.
\begin{table*}
\caption{\label{tab:scaling} Table of the scaling function  $f(z)$}
\begin{tabular}{ |c||c|c|c|c|c|c|c|c|c|c|c|c|c|c|c|c|c|c|c|c|}
\hline

z & 0.1 & 0.2 & 0.3 & 0.4& 0.5& 0.6&0.7&0.8&0.9&1.0&1.1&1.2&1.3&1.4&1.5&1.6&1.7&1.8&1.9&2.0 \\
\hline
f(z) 
&0.01&0.21&0.45&0.65&0.77&0.82&0.82&0.78&0.72&0.64&0.56&0.49&0.42&0.35&0.29&0.24&0.20&0.15&0.12&0.1\\ 
\hline
\end{tabular}
\end{table*}

In Figure~1, we compare our results recorded in Table~II (black-solid curve) with the H1 data for DIS~\cite{DESY} (red data) and the old ALEPH data for $Z$ decay~\cite{ALEPH:1995qic} (grey data).  The agreement is very good for both data sets, supporting the universality of our results.  For comparison we also show the exact asymptotic (\ref{PBXX}) (blue-solid curve), and the KNO particle multiplicity $e^{-z}$~\cite{DIPOLES}, following from the dimensional reduction (diffusive approximation)  of  Mueller$^\prime$s dipole wave-function evolution (dashed-green curve). 
\begin{figure}[!htb]
\includegraphics[width=7cm]{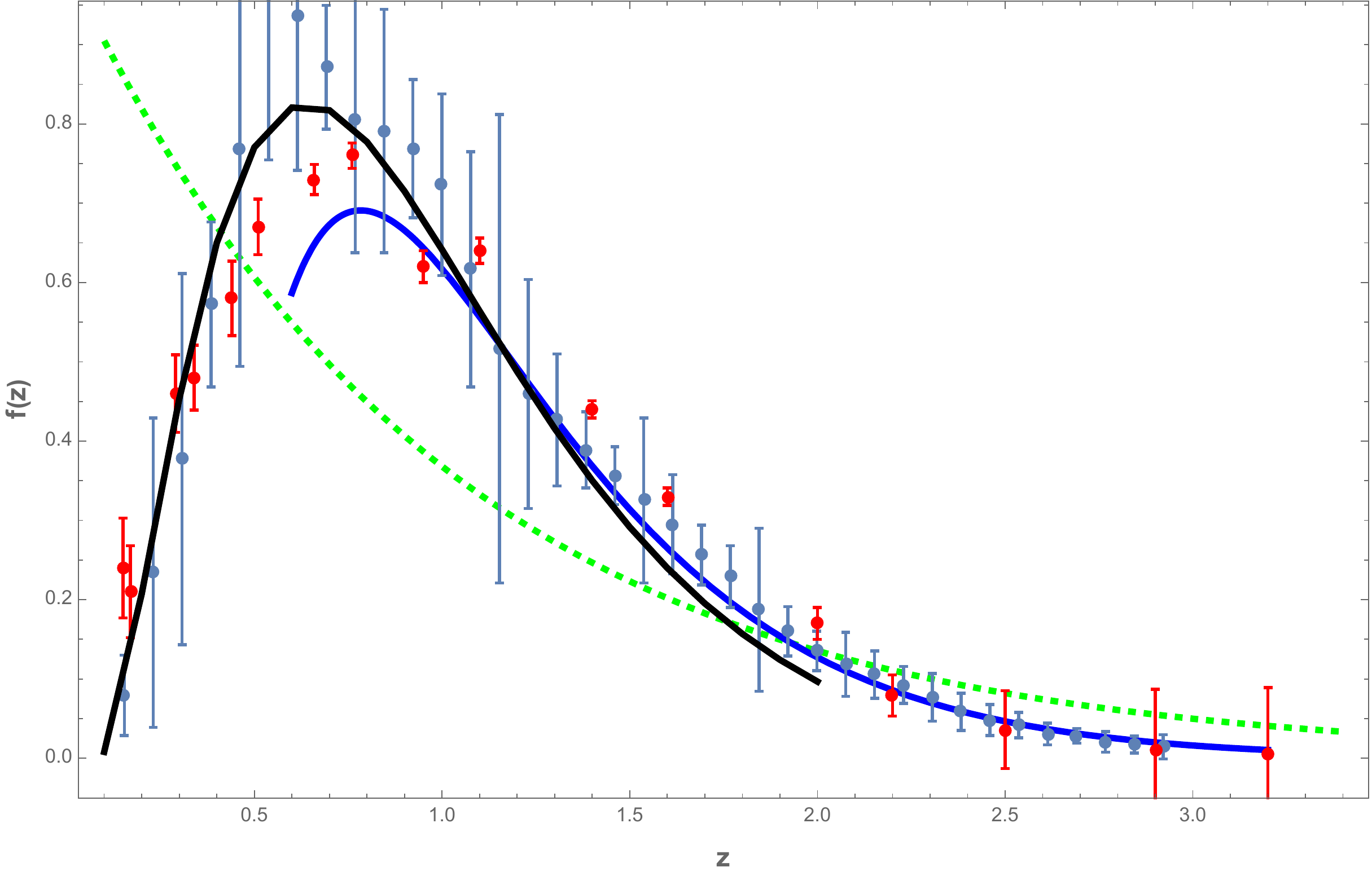}
 \caption{The exact (black-solid curve)  and asymptotic (blue-solid curve) scaling of the KNO particle multiplicity $f(z)$ based on Table II, compared to the recent data for DIS~\cite{DESY} (red) and the $Z$-decay data at $\sqrt{s}=M_Z$ by ALEPH~\cite{ALEPH:1995qic} (grey).}
  \label{fig:asymp}
\end{figure}
% We start  from the observation of Marchesini and Mueller (hep-ph arXiv: 1510.08763v1), that Banfi-Marchesini-Smye equation has a very strong similarity to BK  equation, despite they seem to describe completely different phenomena. The hidden, common feature is the underlying stochastic branching of one object into two objects.  
%\item BMS is a single logarithm equation, so what would  happen if we impose the DLA on soft particle distributions in the jet? Then strong angular ordering appears in the jets, leading to Markov chain of independent "one gluon" into "two gluons" radiations.
%The resulting equation is of Dokshitzer-Fadin-Khoze type equation (5.67) in the book. 
%\item QCD evolution in the dipole picture, despite it has quite a different kernel comparing to BMS,   manifests a similar pattern - if we restrict the splitting of the dipoles to the {\it smaller} dipoles only [Gavin Salam, hep-ph/9504284], then the successive dipole splittings are strongly ordered in size, leading to double logarithm approximation [see also Liou,Mueller,Munier, hep/ph arXiv:1608.00852v1] .
%\item The splitting mechanism is of Markov type and is insensitive to the details of the kernels and/or the fact of  running or not running coupling constant, and again, is given by the identical to (5.67) equation for the generating function. 
%\end{itemize}
\\
\\
\noindent
{\bf 6.}  The knowledge of the effective reduced density matrix for the virtual dipoles in the LFWF~\cite{OUR2}, and the shape of the KNO function, allow  for the evaluation of the entanglement entropy between fast and slow degrees of freedom in DIS~\cite{Stoffers:2012mn,Kharzeev:2017qzs,Kharzeev:2021yyf,OUR2}. For DIS in the DLA with KNO scaling, the result is~\cite{Liu:2022bru}
\begin{eqnarray} \label{eq:entropy}
S_{DIS}(y,Q^2)\rightarrow  {\rm ln}\bar{n} \equiv
2\bigg(\frac{2C_F}{\pi \beta_0}y \ln \ln \frac{Q^2}{M^2}\bigg)^{\frac 12} \ .
\end{eqnarray}
a measure of the Sudakhov contribution. 
(\ref{eq:entropy}) is measurable in the DGLAP regime of DIS.
We note that the KNO scaling function in the diffusive (BFKL)  regime with $f(z)=e^{-z}$~\cite{DIPOLES}, leads to maximal decoherence in the entanglement entropy $S_{BFKL}\sim  y$~\cite{Stoffers:2012mn,Kharzeev:2017qzs,Kharzeev:2021yyf,OUR2}. In contrast, the unimodal character of  the scaling function in the DLA, leads to a smaller  entanglement entropy $S_{DIS}\sim  \sqrt{y}$.

Using the BMS-BK duality, we can readily  formulate the entanglement entropy between soft and hard degrees of freedom in the final state of $e^+e^-$ annihilation into hadronic jets  
\begin{align}
\label{SEEX}
S_{e^+e^-}(Q^2)\equiv
2\bigg(\frac{C_F}{\pi \beta_0}\ln \frac{Q^2}{M^2} \ln \ln \frac{Q^2}{M^2}\bigg)^{\frac 12} \ .
\end{align}
also a measure of  the Sudakov contribution (with no extra 2 in the bracket).
The  rapidity gap between the quark-antiquark pair $y \rightarrow \ln \frac{Q^2}{M^2}$, produces   {\it another logarithm} in $Q^2$. Note that for $e^+e^-$, this contribution dominates the 
single logarithm resummation in the BFKL contribution at large $Q^2$, which is
\bea\ln \bar{n}_{\rm BFKL}\sim \frac{2C_F\ln 2}{\pi \beta_0} \ln \ln \frac{Q^2}{M^2}
\eea
The prediction (\ref{SEEX}) is amenable to experimental verification in high energy hadronic jet physics. 
%Finally, in the last part of this Letter, we discuss the entanglement. In previous paper~\cite{KNOlarge}, we have provided the scheme how to address the entanglement in rapidity between slow and fast degrees of freedom in Mueller's picture. Applying this scheme to DLA approximation adopted  here, we  see that entropy of entanglement grows as a square root of rapidity, in comparison to the diffusion regime (BFKL), when it grows like rapidity. So clearly, not the KNO scaling itself, but the precise shape of the KNO scaling, encodes the quantum information.
%{\bf Taking into account the universality of scaling functions, it is crucial to ask what is the interpretation of quantum entanglement  in jets - what is entangled?, what corresponds to saturation ( hadronization?), are there relations to black hole scrambling ?}
%\begin{acknowledgements}\\
\\
\\
{\bf Acknowledgements:}
\\
We are grateful to Jacek Wosiek for bringing~\cite{OLDJETKNO2} to our attention. This work is supported by the Office of Science, U.S. Department of Energy under Contract No. DE-FG-88ER40388, and by the Priority Research Areas SciMat and DigiWorld under program Excellence Initiative - Research University at the Jagiellonian University. 
%\end{acknowledgements}

\nocite{*}

\bibliography{KNOSHORTBIB}% Produces the bibliography via BibTeX.

\end{document}